\documentstyle[aps,tighten]{revtex}

\begin{document}
\title{Spectroscopy of doubly charmed baryons}
\author{J. Vijande$^{\, 1}$, H. Garcilazo$^{\, 2}$, 
A. Valcarce$^{\, 1}$ and F. Fern\'andez$^{\, 1}$}
\address{$^{\, 1}$ Grupo de F\' \i sica Nuclear \\
Universidad de Salamanca, E-37008 Salamanca, Spain}
\address{$^{\, 2}$ Escuela Superior de F\' \i sica y Matem\'aticas \\
Instituto Polit\'ecnico Nacional, Edificio 9,\\
07738 M\'exico D.F., Mexico}
\maketitle

\begin{abstract}
We study the mass spectrum of baryons with two and three
charmed quarks. For double charm baryons the spin splitting 
is found to be smaller than standard quark-model potential
predictions. This splitting is not influenced either by the particular form
of the confining potential or by the regularization taken for the
contact term of the spin-spin potential. We consistently 
predict the spectra for triply charmed baryons. 
\end{abstract}

\vspace*{2cm} 
\noindent Pacs: 14.20.Lq, 12.40.Yx, 12.39.Jh

\newpage

\section{Introduction}

It has been recently reported the first observation of a candidate for
a double charm baryon, $\Xi^{+}_{cc}$, in data from SELEX \cite{mat02},
the charm hadroproduction experiment at Fermilab. The data are compatible
with a narrow state with a mass of 3520 MeV/c$^2$ decaying through a weak
Cabibbo-allowed process into $\Lambda_c^+ K^- \pi^+$.
This observation has been confirmed through the measurement
of a different weak decay mode that also involves a final state
with baryon number and charm, 
$\Xi^{+}_{cc} \rightarrow p D^+ K^-$ \cite{och04}.
This production region had not been probed by other experiments and 
a big effort is been doing by FOCUS and BELLE looking for 
doubly charmed baryons. SELEX data \cite{moi04} suggest the 
existence of four $ccq$ states ($q$ being a light quark) 
in a mass region of 350 MeV.

The doubly and triply charmed baryons provide a new window for
understanding the structure of all baryons. As pointed out
by Bjorken \cite{bjo85} one should strive to study these systems
because their excitation spectrum should be close to the
perturbative regime. For their size scales the quark-gluon coupling 
constant is small and therefore the leading term in the 
perturbative expansion is enough to describe the system. 
Moreover, the spectroscopy of baryons containing two heavy quarks is of
interest because of similarities both to a quarkonium state, $Q\overline{Q}$,
and to a heavy-light meson, $\overline{Q} q$. On the one hand, the slow
relative motion of the tightly bound
color antitriplet $(cc)_{\bar{3}}$ diquark in $ccq$ 
is similar to quarkonium. On the other hand, for $ccq$
the radius is dominated
by the low mass $q$ orbiting the tightly bound $cc$ pair, and
therefore is large. As a consequence, the relative $(cc)-(q)$
structure may be described similar to $\overline{Q} q$ mesons, where
the $cc$ pair plays the role of the heavy antiquark. The study
of such configurations can help to set
constraints on models of quark-quark forces \cite{fle89,ros95}.
For example, Ref. \cite{sto95} emphasized how the $QQq$ excitation spectra
can be used to phenomenologically determine the $QQ$ potential,
to complement the approach taken for $Q\overline{Q}$ quarkonium
interactions.

Heavy-quark baryons are ideal systems to probe QCD dynamics in close
connection to the structure of heavy-light mesons and the general structure
of hadronic systems. While from the point of view of the 
interacting potential the analysis of light hadronic systems 
becomes complicated by nonperturbative effects, 
heavy-quark systems are rather simple. Heavy-quark current masses
are clear signals of the explicit breaking of chiral symmetry
and as a consequence there are no Goldstone-boson exchange
contributions, the interacting potential being controlled 
by the perturbative short-ranged one-gluon exchange (OGE) and confinement.
From a theoretical point of view, 
in the doubly charmed system one expects a $J=1/2$ ground state 
isodoublet, termed $\Xi^{+,++}_{cc}$ in PDG notation. The $cc$ 
color antitriplet diquark has spin one. The spin of the third quark 
is either parallel, $J=3/2$, or anti-parallel, $J=1/2$, to the diquark.
The $J=3/2$ state has been predicted to be heavier than the $J=1/2$ state by 
around 80 MeV/$c^2$ \cite{fle89,ros95,sto95,ron95,ito00}.
For the $ccc$ system the Pauli principle demands a $J=3/2$ ground state.

Having in mind that the role of models in QCD is to build 
the simplest physical 
picture that connects the phenomenological regularities 
with the underlying structure, in this article 
we use a model that describes correctly the
light baryon and the heavy and heavy-light meson spectra
to study the spectra of doubly and triply charmed baryons.
This procedure, that will have important consequences as we will see in the
following, allows to make parameter-free predictions for the masses of 
doubly and triply charmed baryons. 
For this purpose we first make use of the model of Ref. \cite{vijb4}
designed to describe the meson spectra from the light pseudoscalar mesons
to bottomonium. The model is based on the assumption that the light
quark constituent mass is a consequence of the 
spontaneous breaking of chiral symmetry generating Goldstone
boson exchanges between light quarks. Besides it contains a confining term
and a minimal one-gluon exchange potential.
As mentioned above, for doubly and triply 
charmed baryons chiral symmetry is explicitly
broken and therefore the interacting potential gets reduced to the 
one-gluon exchange and confinement. Let us revise the most 
important aspects of these two contributions.

Following de R\'{u}jula {\it et al.} \cite{ruj75}
the OGE is a standard color Fermi-Breit interaction 
containing a coulomb term plus a spin-spin interaction:
\begin{equation}
V_{OGE}(\vec{r}_{ij}) ={\frac{1}{4}}\alpha _{s}\,\vec{\lambda ^{c}}%
_{i}\cdot \vec{\lambda^{c}}_{j}\,\left\{ {\frac{1}{r_{ij}}}-{\frac{1}{%
6m_{i}m_{j}}}\vec{\sigma}_{i}\cdot \vec{\sigma}_{j}
\,{\frac{{e^{-r_{ij}/r_{0}(\mu )}}}{r_{ij}\,
r_0^2(\mu)}}\right\} \, ,
\end{equation}
where $\vec{\lambda^{c}}$ are the $SU(3)$ color matrices,
$r_{ij}$ is the interquark distance, $m_{i}$ the 
constituent mass of quark $i$, 
$\vec{\sigma}$ are the spin Pauli matrices, 
and $\alpha_s$ is the quark-gluon coupling constant.
The nonrelativistic reduction of the OGE diagram in QCD for point-like
quarks presents a contact term that, when not treated perturbatively,
leads to collapse \cite{bha80}. This is why 
the structure of the OGE is maintained but the $\delta$ function is
regularized in a suitable way. Such regularization takes into account
the finite size of the constituent quarks and should be therefore
flavor dependent \cite{wei83} $r_0(\mu)= r_0 \cdot (\mu_q/\mu_{q_1 q_2})$, 
where $\mu_q$ is the reduced mass of two light quarks and
$\mu_{q_1 q_2}$ is the reduced mass of the two quarks under consideration.
The typical size of the system scales with its reduced 
mass as expected for a coulombic system.

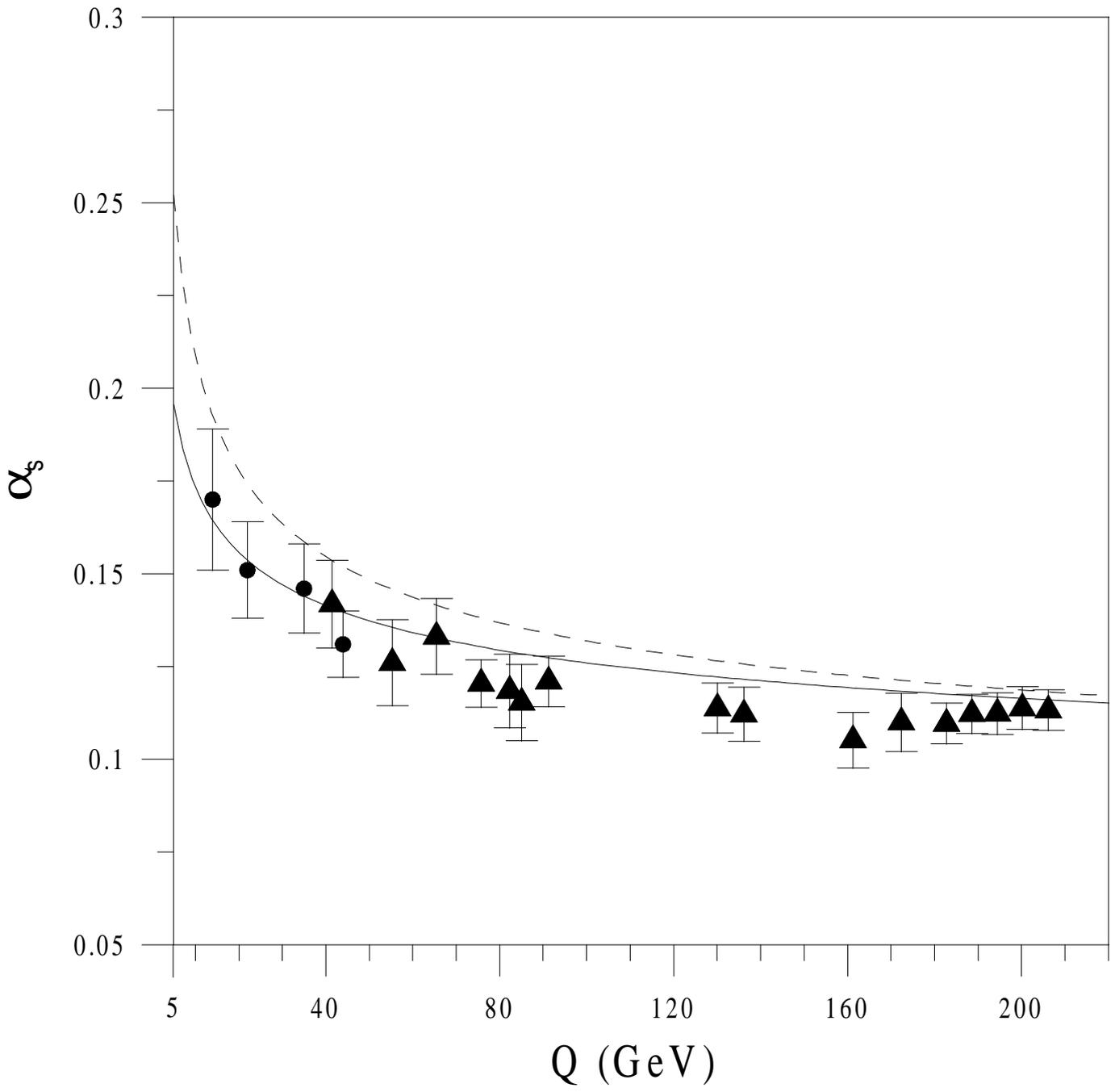
\begin{figure}
\caption{Effective scale-dependent strong coupling constant $\alpha_s$
as a function of momentum.
We plot by the solid line our parametrization.
Dots and triangles are the experimental results
of Refs. \protect\cite{klu03} and \protect\cite{eme02}, respectively.
For comparison we plot by a dashed line the parametrization obtained
in Ref. \protect\cite{shi97} using $\Lambda=$ 0.2 GeV.}
\label{fig1}
\end{figure}

The wide energy covered to describe hadrons made of light and
heavy quarks requires an effective scale-dependent
strong coupling constant \cite{vijb4,tit95} that
cannot be obtained from the usual one-loop expression
of the running coupling constant because it diverges
when $Q\rightarrow\Lambda_{QCD}$. The freezing of the
strong coupling constant at low energies
studied in several theoretical approaches \cite{shi97,bad97}
has been used in different phenomenological models \cite{mat94}.
The momentum-dependent quark-gluon coupling constant is frozen for each
flavor sector. For this purpose one has to determine
the typical momentum scale of each flavor sector that,
as explained in Ref. \cite{hal93}, can be assimilated to the
reduced mass of the system. As a consequence,
we make use of the effective scale-dependent strong coupling constant 
of Ref. \cite {vijb4}, giving rise to the following values
of $\alpha_s$: $\alpha_s (qq)$=0.54, $\alpha_s (qc)$=0.44, and
$\alpha_s (cc)$=0.29. Such scaling generates for the light-quark sector
a value consistent with the one used in the study of the nonstrange
hadron phenomenology \cite{bru02,espe}, and it also has an appropriate
high $Q^2$ behavior, $\alpha_s\sim0.127$ at the $Z_0$ mass \cite{dav97}.
For the sake of consistency we compare in Fig. \ref{fig1} 
the parametrization of Ref. \cite{vijb4} to the experimental
data \cite{klu03,eme02} and the parametrization
obtained in Ref. \cite{shi97} from an analytical model of QCD.

Regarding confinement, lattice calculations in the quenched 
approximation derive, for heavy quarks, a confining 
interaction linearly dependent on the interquark distance,
\begin{equation}
V_{CON}^{L}(\vec{r}_{ij})=  {8 \over 3}  a_{c}\,\,r_{ij} \, .
\label{eqlin}
\end{equation}
This form of strict confinement has been widely
used for light and heavy quarks when studying the meson and baryon 
spectra within a quark model framework. 
The consideration in the lattice of sea quarks apart from valence quarks
(unquenched approximation) suggests a screening effect on the potential when
increasing the interquark distance \cite{bal01}. Creation of light 
$q\overline{q}$ pairs out of vacuum in between the quarks becomes
energetically preferable resulting in a complete screening of quark color
charges at large distances.
In the 80's a specific parametrization of these effects was given in the
form of a screening multiplicative factor in the potential reading $\left[
\left( 1-e^{-\mu r_{ij}}\right) /\mu r_{ij}\right] $ where $\mu$ is
a screening parameter \cite{bor89},
\begin{equation}
V_{CON}^{S}(r_{ij})= {8 \over 3}  a'_c \,\, r_{ij} \left( \frac{
1-e^{-\mu \,\,r_{ij}}}{\mu \,\,r_{ij}}\right) \, .
\label{eqscr}
\end{equation}
Screened confining potentials have been analyzed in the literature 
obtaining significant improvement both for the baryon \cite{vij04}
and for the heavy-meson spectra \cite{vijb4,vij03}.

To get the baryon spectrum we have solved exactly the
Schr\"{o}dinger equation by the Faddeev method.
The Faddeev equations for the case when two particles are identical and the
third is different decouple into two sets corresponding to the two 
possibilities that the wave function be either symmetric or antisymmetric
with respect to the exchange of the two identical ones \cite{afn74,gar90}.
Thus, since the Faddeev formalism includes only the space, spin, and isospin
degrees of freedom one must choose in the case of the three-quark problem
the set which is symmetrical under the exchange of space, spin and isospin,
since the color part of the wave function is already assumed to be 
antisymmetric under the exchange of any two quarks.
In order to assure convergence 
we shall include ($l,\lambda ,s,t$) configurations ($l$ is the orbital
angular momentum of a $2q$ pair, $\lambda $ is the orbital angular momentum
of the third quark with respect to the center of mass of the $2q$, and $s$
and $t$ are the spin and isospin of the $2q$ respectively) up to $l$=5 and 
$\lambda $=5 \cite{espe}. 
For the case of three identical particles we have also
calculated the spectra by means of the hyperspherical
harmonic (HA) expansion method \cite{bar00}.
The HA treatment allows a more intuitive understanding of the wave functions in
terms of the hyperradius of the whole system. As a counterpart one has to go
to a very high order in the expansion to get convergence. To assure this we
shall expand up to $K=24$ ($K$ being the great orbital determining the order
of the expansion). 
Differences in the results for the $3q$ bound state
energies obtained by means of the two methods turn out to be at most of 5 MeV. 

Our results for the excitation energy of the low-lying $ccq$ states 
for the model of Ref. \cite{vijb4}
are presented in Table \ref{tab0} together with 
the corresponding results for the $qqq$ system and 
their experimental values. As can be observed,
the first radial excitation lies
above the first negative parity state.
This is the same
situation found in the light baryon spectra where OGE quark-model based
potentials predict the first negative parity
excitation below the so-called Roper resonance, the first radial
excitation. For double charmed baryons (DCBs) this is, however, 
expected to correspond to the experimental
situation opposite to the light baryon case. The reverse of the ordering
between the positive and negative nucleon excited states has been
explained in terms of the combination of two different effects: 
relativistic kinematics and the pseudoscalar Goldstone-boson
exchange interaction between
light quarks. The nucleon Roper resonance is particularly sensitive
to the form of the kinetic energy operator \cite{car83,gar03}, its 
energy being decreased with respect to the first negative 
parity state when a relativistic kinetic energy operator is used.
Once this energy difference has been diminished, the pseudoscalar 
Goldstone boson exchange potential
produces the desired inversion \cite{gar03,glo96}.
In DCBs kinetic energy relativistic effects
are expected to be much smaller due to the presence of two heavy
quarks that makes the system analogous to a hydrogen-like atom
\cite{fle89}. On the other hand, chiral symmetry is explicitly broken
and the interaction between heavy-light or heavy-heavy quarks
does not present a Goldstone-boson pseudoscalar term. 
These two effects combined
should recover the normal ordering between positive and negative
parity states in the case of doubly charmed baryons.

Another striking result appearing in the obtained spectra concerns
the $J^P=1/2^+ - J^P=3/2^+$ spin splitting, it appears to be 
much smaller than the one predicted based on potential 
models directly obtained from the $c\bar c$ spectra
\cite{fle89,ros95,sto95,ron95,ito00}.
Such a small spin splitting seems to be related with
our scale-dependent quark-gluon coupling constant and
the fact that the $\Delta - N$ mass difference is 
correctly reproduced. 

As has been discussed above the interacting potential 
for doubly and triply charmed
baryons may depend on i) the specific form used for the confining
term, and ii) the regularization taken for the contact term of the
one-gluon exchange potential. To judge the model dependence of our 
results on these items, we have recalculated the baryon spectra
with four different set of parameters that are given in Table \ref{tab1}.
Two of them, sets B and D, have an infinite linear confining potential
as the one of Eq. (\ref{eqlin}), while sets A and C present a screened
confining potential, as the one of Eq. (\ref{eqscr}). 
For both types of confining potentials we have used two different
values of the regularization parameter for the spin-spin interaction, $r_0$.
In all cases the $\Delta - N$ mass difference is 
asked to be correctly described. 

The results for the $ccq$ system
are presented in Table \ref{tab3} and the corresponding
ones for the $qqq$ system in Table \ref{tab2}.
As can be seen on these tables the same conclusions as before hold for any
of the set of parameters used. All set of parameters
give comparable radial and orbital excitation energies, 
the relative position of the
first positive and negative parity excited states being preserved.
Besides the spin-splitting appears to be almost constant (see
the third column of Table \ref{tab3}) independently of the set
of parameters used and much smaller than the one predicted 
on the existing literature (one should have in mind at this 
point that most existing calculations are of rather exploratory
nature, since made when double charm physics was considered far
future).
This result is independent of the type of confinement used, screened 
or infinite, and also of the strength of the spin-spin force controlled 
by the values of $\alpha_s$ and $r_0$, as it is illustrated in 
Table \ref{tab4}. Such a small spin splitting resembles the situation
with the $\Upsilon -\eta_b(1S)$ mass difference, being very difficult
to disentangle from the experimental point of view, but it should
stress efforts in order to try to find evidence for the existence
of two peaks in the same energy region.
It also does not appear affected by the form of the confining interaction,
in agreement with conclusions of Ref. \cite{vij04}, that observes 
that only the high-energy excited states seem to be ''confinement 
states'', and therefore influenced by the explicit form of the confining
potential.

The mass of the $ccq$ $J^P=1/2^+$ ground state can be tuned to reproduce
the measured experimental data by means of the constituent charm quark mass. 
This has been done by means of the set of parameters E in Table \ref{tab1}, 
that is exactly the same as the set A except for 
the constituent charm quark mass (the same could have been done for
any of the other sets of Table \ref{tab1}). The results are shown
in Table \ref{tab3new}. As can be seen there are no 
variations in the predicted structure of the excited states.
The spin splitting is exactly the same as before
as well as the general structure of the spectra that remains 
also independent on modifications of the charm constituent 
quark mass, the excitation energies being almost the 
same as those presented in Table \ref{tab3} for the set of 
parameters A, but reproducing in this case the experimental
energy of the double charm baryon, $\Xi^+_{cc}$, reported
by SELEX, 3520 MeV/c$^2$ \cite{mat02,och04}.

Using the set of parameters E we have also calculated the 
spectra for the $ccc$ system. The results are given in Table \ref{tab5}.
In this case it is the $J^P=3/2^+$ the lighter state. As in the $ccq$ system
the first negative parity state appears below the first radial excitation.
The spin splitting increases up 
to values close to the results obtained for the light-quark baryons. 
Therefore, the small spin-splitting found in this work
is a characteristic feature of double-heavy baryons.

As a summary, we have predicted the $ccq$ and $ccc$ spectra by means of a
potential model that describes the light-baryon
and the heavy and heavy-light meson spectra. 
The results are tested against different forms of the
confining interaction and different values of the regularization 
parameter of the delta term in the
OGE potential. Two results are specially relevant. First of all, the 
$J^P=1/2^+ - J^P=3/2^+$ spin-splitting is obtained to be almost 
constant independently of the model used and smaller than the results
reported in the past, of the order of 20 MeV. 
It seems to be related with the correct description
of the $\Delta - N$ mass difference. Secondly, the normal ordering between the 
positive and negative parity excited states is recovered, what 
should be the consequence of 
the absence of pseudoscalar forces and the reduced influence
of relativistic effects in the DCBs spectra.
The increasing interest and the actual experimental possibilities
in the charm sector claims for an 
experimental effort to disentangle these questions.

\acknowledgements
This work was partially funded by Direcci\'{o}n General de
Investigaci\'{o}n Cient\'{\i}fica y T\'{e}cnica (DGICYT) under the Contract
No. BFM2001-3563, by Junta de Castilla y Le\'{o}n under the Contract No.
SA-104/04, and by COFAA-IPN (Mexico).

\begin{table}[tbp]
\caption{Relative energy $ccq$ and $qqq$ spectra, in MeV, for the 
model of Ref. \protect\cite{vijb4}. For the $qqq$ system 'Theor.'
stands for the results of the model of Ref. \protect\cite{vijb4}
and 'Exp.' for the experimental result.}
\label{tab0}
\begin{tabular}{cccccc}
& System  & $J^P(1/2^+)^*$ & $J^P(3/2^+)$ & $J^P(1/2^-)$ &  \\ 
\hline
&   $ccq$             & 287    &  25   &  206  &  \\ 
&   $qqq$ (Theor.)     & 500    &  290  &  469  &  \\ 
&   $qqq$ (Exp.)       & 491$-$531    &  291$-$295  &  581$-$616  &  \\ 
\end{tabular}
\end{table}

\begin{table}[tbp]
\caption{Different parameter sets used in the text.}
\label{tab1}
\begin{tabular}{ccccccc}
&       & A & B & C & D & E\\
\hline 
& $m_q$ (MeV)            &\multicolumn{4}{c}{313}    & 313 \\
& $m_c$ (MeV)            &\multicolumn{4}{c}{1752}   & 1550  \\
& $r_0$ (fm)             &\multicolumn{2}{c}{0.25}&\multicolumn{2}{c}{0.45} & 0.25  \\ 
\hline
Linear confinement & $a_c$ (MeV fm$^{-1}$)  & $-$  &  55  & $-$  & 115   & $-$ \\
Screened confinement & 
$\left\{ \begin{array}{c} a'_c ({\rm MeV}) \\ \mu ({\rm fm}^{-1}) \end{array} \right. $ &
$\begin{array}{c} 160 \\ 0.7 \end{array}$ &
$\begin{array}{c} - \\ - \end{array}$ &
$\begin{array}{c} 300 \\ 0.7 \end{array}$ &
$\begin{array}{c} - \\ - \end{array}$ & 
$\begin{array}{c} 160 \\ 0.7 \end{array}$  \\
\end{tabular}
\end{table}

\begin{table}[tbp]
\caption{Relative energy $ccq$ spectra, in MeV, for the first four sets
of parameters of Table \protect\ref{tab1}.}
\label{tab3}
\begin{tabular}{cccccc}
& Set    & $J^P(1/2^+)^*$ & $J^P(3/2^+)$ & $J^P(1/2^-)$ &  \\ 
\hline
&   A    & 221    &  23   &  165  &  \\ 
&   B    & 221    &  22   &  156  &  \\ 
&   C    & 347    &  25   &  243  &  \\ 
&   D    & 332    &  21   &  217   &  \\ 
\end{tabular}
\end{table}

\begin{table}[tbp]
\caption{Relative energy $qqq$ spectra, in MeV, for the first four sets
of parameters of Table \protect\ref{tab1}.}
\label{tab2}
\begin{tabular}{cccccc}
& Set    & $J^P(1/2^+)^*$ & $J^P(3/2^+)$ & $J^P(1/2^-)$ &  \\ 
\hline
&   A    &  408    &  291  &  403  &  \\ 
&   B    &  473    &  292  &  430  &  \\ 
&   C    &  591    &  299  &  548  &  \\ 
&   D    &  689    &  297  &  580  &  \\ 
\end{tabular}
\end{table}

\begin{table}[tbp]
\caption{Strength of the spin-spin force, $\alpha_s / r_0^2(\mu)$ in fm$^{-2}$,
for the different quark-quark pairs and for the different sets of parameters
 of Table \protect\ref{tab1}.}
\label{tab4}
\begin{tabular}{ccccc}
& & Sets A and B & Sets C and D &  Set E \\ 
& {\rm Quark pair} & $r_0=0.25$ & $r_0=0.45$ & $r_0=0.25$  \\ 
\hline
& $qq$ &  8.64        &  2.66    &  8.64 \\ 
& $qc$ &  20.36       &  6.26    &  19.11 \\ 
& $cc$ &  149.79      &  45.31   &  119.2 \\ 
\end{tabular}
\end{table}

\begin{table}[tbp]
\caption{$ccq$ spectra, in MeV, for the set of parameters 
E of Table \protect\ref{tab1}.}
\label{tab3new}
\begin{tabular}{cccccc}
& $J^P(1/2^+)$ & $J^P(1/2^+)^*$ & $J^P(3/2^+)$ & $J^P(1/2^-)$ &  \\ 
\hline
& 3524        &  3749 (225)    &  3548 (24)   &  3692 (168)  &  \\ 
\end{tabular}
\end{table}

\begin{table}[tbp]
\caption{$ccc$ spectra, in MeV, for the set of parameters 
E of Table \protect\ref{tab1}.}
\label{tab5}
\begin{tabular}{cccccc}
& $J^P(3/2^+)$ & $J^P(3/2^+)^*$ & $J^P(1/2^+)$ & $J^P(1/2^-)$ &  \\ 
\hline
&  4632         &   4870 (238)   &  4915 (283)  &  4808 (176)  &  \\ 
\end{tabular}
\end{table}

\end{document}